# THE DWARF NOVA PQ ANDROMEDAE

JOSEPH PATTERSON,[1] JOHN R. THORSTENSEN,[2]

EVE ARMSTRONG,[3,1] ARNE A. HENDEN,[4,5] AND ROBERT I. HYNES[6]

---

[1] Department of Astronomy, Columbia University, 550 West 120th Street, New York, NY 10027; jop@astro.columbia.edu

[2] Department of Physics and Astronomy, Dartmouth College, 6127 Wilder Laboratory, Hanover, NH  03755; j.thorstensen@dartmouth.edu

[3] Department of Physics, University of California at San Diego, Mail Code 0354, 9500 Gilman Drive, San Diego, CA  92093; earmstrong@physics.ucsd.edu

[4] Universities Space Research Association, United States Naval Observatory, Flagstaff Station, Post Office Box 1149, Flagstaff, AZ  86002; aah@nofs.navy.mil

[5] American Association of Variable Star Observers, 25 Birch Street, Cambridge, MA  02138; arne@aavso.org

[6] Department of Physics and Astronomy, Louisiana State University, Baton Rouge, LA 70803; rih@phys.lsu.edu

## ABSTRACT

We report a photometric study of the WZ Sagittae-type dwarf nova PQ Andromedae. The light curve shows strong (0.05 mag full amplitude) signals with periods of 1263(1) and 634(1) s, and a likely double-humped signal with $P$=80.6(2) min. We interpret the first two as nonradial pulsation periods of the underlying white dwarf, and the last as the orbital period of the underlying binary. We estimate a distance of 150(50) pc from proper motions and the two standard candles available: the white dwarf and the dwarf-nova outburst. At this distance, the $K$ magnitude implies that the secondary is probably fainter than any star on the main sequence — indicating a mass below the Kumar limit at 0.075 $M_\odot$. PQ And may be another "period bouncer", where evolution now drives the binary out to longer period.

*Subject headings*: accretion, accretion disks — binaries: close — novae, cataclysmic variables — stars: individual (PQ Andromedae)

## 1. INTRODUCTION

The oldest cataclysmic variables should be binaries of short orbital period, in which the accretion rate is very low and the secondary has been whittled down to 0.02–0.05 $M_\odot$. This is the natural end-state predicted by the simplest theories of binary evolution (Paczynski 1981; Rappaport, Joss, & Webbink 1982; and their successors). These oldsters have proved quite elusive, though, because they are very infrequent erupters and because their "substellar" secondaries (i.e. below the Kumar limit at 0.075 $M_\odot$) emit so little light. This class of CV should evolve quite slowly, because the mass ratio $q=M_2/M_1$ is very low and therefore gravitational-radiation losses are very weak (roughly $\dot{J} \sim q^2$), CVs evolve relatively rapidly in their earlier phases, and therefore the oldest ones should dominate a complete census. Kolb (1993) estimated that ~70% of CVs should be of this type, assuming that late evolution is dominated by GR losses. Nevertheless, there is still not wide agreement as to whether the class even exists. At the 2004 Strasbourg conference, Tom Marsh turned to the audience and asked "Does anyone know the *names* of any of these stars?"... and was answered by silence.

This silence testifies to the limited success of our efforts, but we have been trying to identify these very old CVs — sometimes called "period bouncers" since they have in principle evolved past minimum period (Patterson 1998, 2001). Littlefair et al. (2003) and Patterson (2001) discuss the difficulties in certifying period-bouncer status, and in discovery. Recently we presented a list of seven promising candidates: stars of demonstrably very low $q$ (<0.06, Patterson, Thorstensen, & Kemp 2005; hereafter PTK, and with the evidence basically contained in their Figure 9). These are particularly interesting because they appear to be quite nearby — including (formally) the two nearest hydrogen-rich CVs in the sky. So despite their great bashfulness, these stars may indeed comprise the majority species of CVs.

Dwarf novae of very long recurrence period, often called "WZ Sagittae stars", are plausible candidates for this class. Here we present a study of one such star, PQ Andromedae.

## 2. BACKGROUND

PQ And was discovered in 1988 at magnitude 10 by D. McAdam, and declined by 2 mag over the next 19 days (Hurst et al. 1988a, b; Hurst & Young 1988). Several months after outburst, with the remnant at $V\sim18$, the spectrum showed Balmer emission lines nestled in the midst of broad absorptions (Wade & Hamilton 1988). Richter (1990) searched archival plates and identified earlier eruptions in 1938 and 1967, thus establishing a recurrence period of ~25 yr (or less, since some outbursts may have been missed). These photometric and spectroscopic properties are commonly found in the WZ Sge subclass of dwarf novae, and the star is now considered in that class. A recent study by Schwarz et al. (2004) is consistent with this, and suggested an underlying orbital period near 100 min.

## 3. OBSERVATIONS

### *3.1 ASTROMETRY*

In preparation for a parallax study, we obtained a precise J2000.0 astrometric position,

corrected for proper motion, from a 2002 October image:

$$RA = 2^h\ 29^m\ 29.60^s$$
$$Dec = +40°\ 02'\ 39.9"$$

The plate solutions were fit with UCAC–2 stars, and rms residuals were below 0.1". The total error was near 0.1" in each coordinate.

Combining this position with data retrieved and measured from the USNO archives, we obtained a proper motion of

$$\mu_x = +39(3)\ mas/yr$$
$$\mu_y = +22(3)\ mas/yr.$$

The USNO–B archive alone, without the recent precise CCD image, gives a similar answer: $\mu_x=+34(6)$ and $\mu_y=+24(4)$ mas/yr.

This is a large proper motion, suggesting a fairly nearby star. With our favored velocity ellipsoid for CVs (Thorstensen 2003), the mean distance is 83 pc, with the 16th and 84th percentiles of the distribution at 47 pc and 148 pc. However, the distribution is always quite asymmetric (see for example the dashed curve in Figure 2 of Thorstensen 2003), so the most probable distance is ~120 pc.

### 3.2  PERIODICITY SEARCH

In 2005 January we obtained four consecutive nights of time-series photometry, using a CCD imager on the MDM 1.3 m telescope. We used a broadband (3800–8000 Å) filter and several integration times in the range 40–60 s. PQ And showed no obvious variability in its light curve, staying always close to $V=18.5\pm0.4$. (The error was large because the passband was so broad.)

To search for periodic signals, we calculated the power spectrum of the four-night light curve. The result is seen in the upper frame of Figure 1. Two obvious periodic signals are flagged, at 68.41(5) and 136.36(5) c/day. The complex of power near 36 c/d is likely to contain a third signal — probably at 35.70(6) c/d, although the alias at 36.70 c/d is a viable and worthy alternative. The lower frames contain close-ups of these three signals, and synchronous summations. In the case of the signal at lowest frequency, we summed at *half* the detected frequency (i.e. 17.85 c/d) — because there was a weak complex near the 18/19 c/d subharmonic, and because double-humped orbital waveforms are very common in low-$\dot{M}$ dwarf novae. Naturally, the summation at 0.0560 d (17.85 c/d) does show a double-humped waveform. We shall refer to it as the "orbital" waveform, although the hypothesis of an orbital origin is unproven.

### 3.3  CALIBRATED PHOTOMETRY

We observed PQ And several times with calibrated photometry. On the nights of 17

January and 28 January 2004, we observed with the 1.0 m reflector of the U.S. Naval Observatory. We found the star respectively at $V$=19.24(8), $B$=19.25(8) — and at $V$=18.94(10), $B$=19.07(13). This yields a mean of $V$=19.10, $B$–$V$=0.07. Our estimate during January 2005 was slightly brighter, so we adopt $V$=18.95. The total galactic reddening on this line of sight is $E(B–V)$=0.06 (Schlegel, Finkbeiner, & Davis 1998); and for distances near 200 pc, most of this reddening is probably also on the line of sight to PQ And. Assuming $A_v$=3 $E(B–V)$, we adopt $A_v$=0.15 mag and hence a dereddened $V$=18.8.

Infrared photometry was obtained on 2005 January 20 with the Simultaneous Quad Infrared Imaging Device (SQIID) on the 2.1m telescope at Kitt Peak National Observatory. The camera records $J$, $H$, $K$, and $L$ images simultaneously, although we found that the $L$ data were of insufficient quality to be useful. $JHK$ data were all of good quality and obtained synchronously with 5×8s exposures obtained at each of ten dithered positions. The images were combined to produce a sky image which was subtracted from each target image, before combining them into an average for each band. Each of these averages preserved an image quality better than 1.4". Sensitivity variations were corrected with twilight sky flats in the usual manner.

PQ And was adequately detected in all filters. Standard aperture photometry was used to extract differential magnitudes of the target relative to several 2MASS stars in the field, all of which were near 14th magnitude in $J$, $H$, and $K$ — well exposed and with 2MASS errors of less than 0.05 mag. The derived magnitudes of PQ And were $J$=18.57(13), $H$=17.97(11), $K$=17.45(10), where the uncertainties are dominated by statistics of the PQ And measurement itself.

## 4. INTERPRETATION

### 4.1 DISTANCE

WZ Sge stars typically become very faint in quiescence, with the visual light dominated by the white dwarf. The broad Balmer absorptions in PQ And certainly support this. In addition, the noncommensurability of the 68 and 136 c/d photometric signals strongly suggest an origin in nonradial pulsations of the white dwarf — indicating again that the white dwarf dominates the light. We estimate that ~75% of the quiescent V light comes from the white dwarf, with the remainder coming mostly from accretion (because of the likely double-humped orbital signal, and because of the nightly variability). This implies $V$=19.1 for the white dwarf, and $V$=20.3 for accretion light.

The white dwarf's temperature was measured by Schwarz et al. as 12000±1000 K, and a similar temperature is suggested by the presence of white-dwarf pulsations, assuming the white dwarf to be like ordinary ZZ Cet stars ($T_{eff}$=11500±600 K, Mukadam et al. 2004). White dwarfs in CVs average ~0.8 $M_\odot$, which implies $M_v$=12.4 for this $T_{eff}$. Thus we estimate a "white dwarf parallax" of 220(60) pc.

The dwarf-nova outburst itself also provides a distance estimator. Supermaxima of 80-minute SU UMa-type dwarf novae (the parent class of the WZ Sge subclass) should reach $M_v$~4–5 (Warner 1987; Cannizzo 1988; Harrison et al. 2003), which suggests a distance of 140(50) pc.

A third constraint comes from proper motions, which we discuss above and roughly characterize as 110(50) pc. An average of these three constraints implies a final distance estimate of 150(50) pc.

Because PQ And appears so similar to WZ Sge, a star of precisely known distance (43±3 pc, Thorstensen 2003; Harrison et al. 2003), we can also try to use WZ Sge to scale the distances. The dereddened white-dwarf components have $V$=19.1 and 16.2, and the dwarf novae at maximum have $V$=10.0 and 8.0. These suggest that the distance moduli differ by 2.4(5) mag, implying $d$=130(30) pc. A second useful comparison star is GW Lib, another WZ Sge-type dwarf nova — with a similar quiescent spectrum and light curve (including nonradial pulsations, certifying a similar $T_{eff}$ for the white dwarf: van Zyl et al. 2004). Trigonometric parallax yields a distance of 105(20) pc to GW Lib (Thorstensen 2003). The white-dwarf components have $V$=19.1 and 17.7 in PQ And and GW Lib, and the dwarf novae at maximum had $V$=10.0 and 8.5. Comparison suggests $d$=200(60) pc for PQ And.

These are consistent with our estimated 150(50) pc. That's our story, and we're sticking to it.

### *4.2 FLUX DISTRIBUTION, AND LUMINOSITY OF COMPONENTS*

The dereddened flux distribution is shown in Figure 2. The points show the *BVJHK* measurements, and the curve shows the flux distribution of the pulsating white dwarf G226–29. The two flux scales differ by a factor 560, which sets the G226–29 *V* point at that of the estimated white-dwarf component in PQ And (75% of the total). Thus the curve represents our estimate of the full flux distribution of the white dwarf in PQ And, and the difference between points and curve represents other components (secondary star and accretion light).[7] As shown in Figure 2, that difference is roughly constant at ~0.035 mJy from *K* through *B* bands, with possibly a little turnup at *H* and *K*.

Accretion light in such stars is likely to be optically thin emission with $T$~5000–7000 K (Williams 1980, Tylenda 1981), and at low frequency this typically shows a spectral slope $F_\nu \propto \nu^0$. This seems to be the main lesson of Figure 2 also: the second component in PQ And is very likely the accretion disk, with its attendant "hot spot".

Still, the possible *HK* turnup leaves room for some contribution by a cool secondary. After subtracting the *K* contribution of the white dwarf and a minimal estimate for accretion light (extrapolating from *BV* measurements), we estimate from Figure 2 that perhaps half of the *K* light could come from the secondary. This leaves an upper limit of *K*=18.2(3) for the secondary, or $M_K$>12.3(6) at the estimated distance.

This is probably too faint to be a main-sequence star. At the end of the (core hydrogen burning) main sequence at 0.075 $M_\odot$, model stars have $M_K$=11.4 at an age of 10 Gyr, and $M_K$=10.8 at 1 Gyr (Baraffe et al. 1998). It is not obvious which age is appropriate to use for a CV secondary, which is "ageless" since it does not evolve on a single-star timescale. However, secondaries burning hydrogen in their cores must be *at least* as bright as a "ZAMS" model (10 Gyr for low-mass stars), because the quirks of CV evolution (mainly the lingering effects of energy deposition in the envelope from prior evolution and heating episodes) can only add to this luminosity. Thus we conclude that the secondary is essentially "substellar", i.e. with $M_2$<0.075 $M_\odot$.

---

[7] Of course, the accuracy of this subtraction depends on the assumption that the white dwarf in PQ And has a temperature similar to that of ZZ Cet stars. Further study on this point is very desirable.

The accretion light is also very faint, at $M_v$=14.4(9). This is among the faintest accretion components seen among CVs, and further emphasizes the likelihood that PQ And is a period-bouncer — an old CV which has evolved past minimum period and cannibalized its secondary down to $M_2$<0.075 $M_\odot$ (PTK, esp. their Table 4).

### 4.3 COMPARISON WITH SCHWARZ ET AL.

Our estimates of distance and (therefore) luminosity are in conflict with those given by Schwarz et al. (2004), who fit the absorption-line profiles to a model white dwarf and found $T_{\text{eff}}$=12000 K, log $g$=7.7, $d$=330(50) pc. We are inclined to distrust such fits for white dwarfs in CVs; $T_{\text{eff}}$ is straightforward if the white dwarf dominates the flux, but log $g$ depends on high-quality spectra in the line wings, plus the assumption that line wings are uncontaminated by emission components (or a way to identify and subtract those components). In particular, the three Balmer lines used by Schwarz et al. are all threatened by He I emission lines in their wings, plus the broad base of Balmer emission (unmeasurable, but still in principle contaminating the profiles). And judging from the strength of the He I 5876 emission in Figure 1 of Schwarz et al., we reckon that the weaker He I lines are significant contaminants. Fill-in of the absorption would lower the apparent log $g$. An undesired and unsubtracted continuum flux (from accretion, which we reckon at ~25% from the putative orbital signal) would also bias the result toward lower gravity. So we suspect that Schwarz et al. deduced too low a gravity, and prefer to merely assign an $M_v$ appropriate for a nominal 0.8 $M_\odot$ (Webbink 1990, Smith & Dhillon 1998) white dwarf, along with the 12000 K reasonably certified by the pulsations. The latter also yields a distance much more compatible with other methods of estimation (proper motion and the dwarf-nova outburst).

Schwarz et al. quote another distance estimate even more discrepant from ours: a lower limit of 319–817 pc from the observed $K$ flux. This is from application of the "Bailey method": assuming that the secondary is of known surface brightness, in this case appropriate for a M5–M7 secondary ($S_K$=5.5±0.5, Ramseyer 1994). We have no quarrel with Bailey's method, but the very short orbital period that we favor (80 or 78 min, compared to ~100 min in Schwarz et al.) greatly changes the resultant distance estimate.

Bailey's method uses the secondary's radius $R_2$, surface brightness $S_K$, and $K$ magnitude to yield the distance $d$ in pc:

$$S_K = K - 5 \log (d/10) + 5 \log (R_2/R_\odot).$$

PQ And is essentially at the minimum $P_{orb}$ for hydrogen-rich secondaries, implying a secondary at or beyond the end of the main sequence. The value of $S_K$ for such stars is unknown, and probably differs very greatly if the star has ceased nuclear burning. In short, the Bailey method is uncalibrated in this regime. In WZ Sge, for example, where all parameters on the right side of the equation are specified by observation ($K>16$, $d=43$, $R_2=0.09$), we obtain $S_K>7.8$. If we adopt $K=18.2$ and apply this to PQ And (not unreasonable since the two stars are apparently so similar in $P_{orb}$), we would obtain $d<125$ pc.

## 5. SUMMARY

1. We report noncommensurate periods of 1263 and 634 s in the time-series photometry of PQ And at quiescence. These probably originate from nonradial pulsation of the underlying white dwarf. The white dwarf appears to contain ~75% of the $V$ flux, or $V=19.1(2)$.

2. Another signal appears to be present at 35.70 or 36.70 c/d. Since double-humped orbital variations are very common in WZ Sge stars, we tentatively interpret this as the signature of an orbital period of 80.7 or 78.5 (±0.2) min. This needs confirmation, preferably with a radial-velocity study.

3. Three methods of distance measurement converge on an estimate of 150(50) pc. This assigns $M_v=13.0(7)$ to the white dwarf, and $M_v=14.4(10)$ to accretion light. The accretion component is extremely faint — apparently fainter than WZ Sge, which has long been the "poster child" for intrinsically faint CVs.

4. *BV* measurements and the presence of pulsations suggest a white-dwarf temperature near 12000 K, where ZZ Cet stars live. *JHK* measurements show red colors ($J-K=1.1$), indicating a cooler source. This could be the secondary star, or cool optically thin emission from the disk. Both may be significant contributors in the infrared, but the disk probably dominates. Formally, we estimate contributions to the $K=17.45(10)$ flux as follows: $K=19.4(2)$ from the white dwarf, $K=18.5(5)$ from accretion light, and 18.3(4) from the secondary. The secondary could, however, be much fainter. These estimates could be refined with an infrared spectrum of good quality, or a more thorough study of the flux distribution (especially the accretion light, which we have not been able to study well since it does not provably dominate any of the bands observed).

5. At 150 pc, this deconvolution implies $M_K=12.4(5)$ for the secondary. This is fainter by 1–1.5 mag than the faintest stars on the hydrogen-burning main sequence, and the secondary is therefore likely to have a mass below the Kumar limit at 0.075 $M_\odot$. This is a characteristic of the "period bouncers" among the dwarf novae.

J.P. acknowledges financial support from NSF grants AST–0098254 and –0406813, and J.R.T. acknowledges support from NSF grants AST–9987334 and –0307413. R.H. gratefully acknowledges the assistance of Kevin Pearson and Dawn Gelino in obtaining the infrared photometry. This publication makes use of data products from the Two Micron All Sky Survey, which is a joint project of the University of Massachusetts and the Infrared Processing and Analysis Center/California Institute of Technology, funded by NASA and the NSF. It has also made use of the USNOFS Image and Catalogue Archive operated by the U.S. Naval Observatory, Flagstaff Station.

# FIGURE CAPTIONS

FIGURE 1. — *Upper frame*, power spectrum of the four-night time series, with two strong signals flagged with their frequency in cycles/day (±0.05). In addition there is a probable signal near 36 c/d. *Other frames:* magnified views of the relevant frequency regime, and folded waveforms at right. The "orbital" signal is folded on half the candidate 35.70 c/d signal.

FIGURE 2. — The points represent the *BVJHK* fluxes of PQ And at quiescence. The curve is the known flux distribution of the pulsating white dwarf G226–29, which should resemble the white dwarf in PQ And if the temperatures are similar. The G226–29 *V* flux is scaled to appear at 75% of the *V* flux of PQ And (the estimated white-dwarf contribution). Thus the *excess* above the curve represents light from other sources in the binary, mostly radiation from the disk. The secondary might account for a possible small turnup in the *K* band.

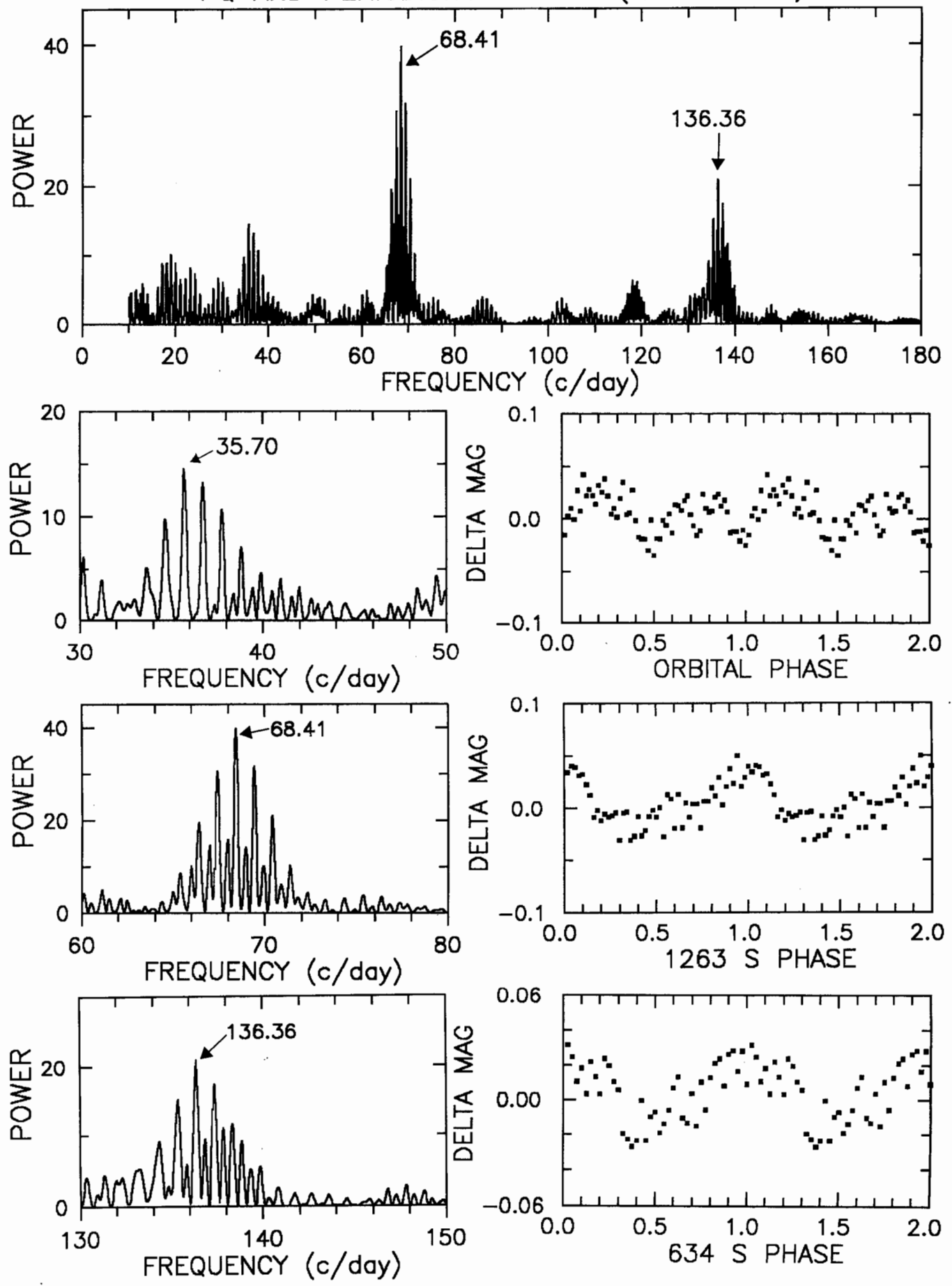

Fig 1

PQ AND PERIODICITY SEARCH (QUIESCENCE)

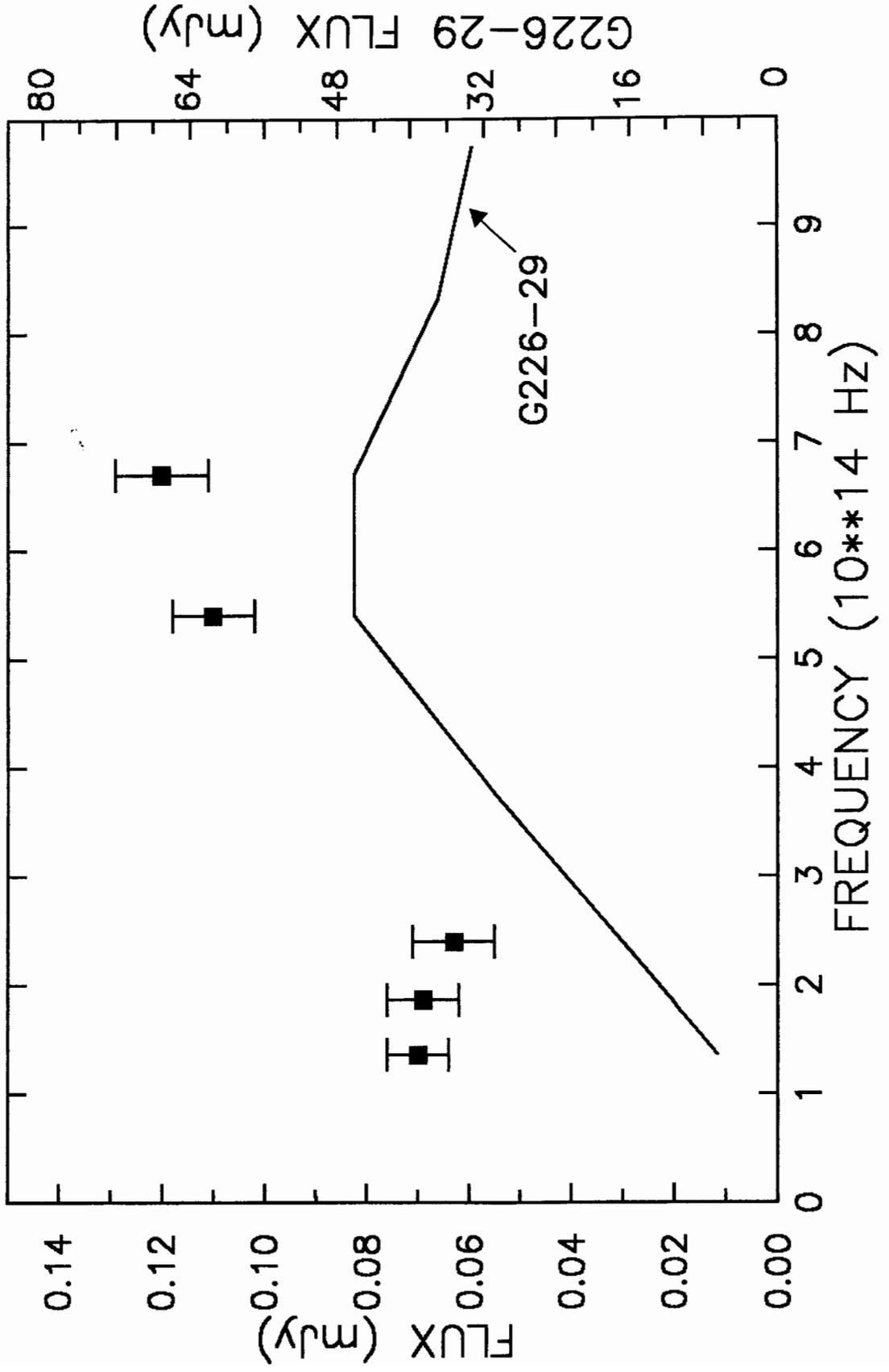

Fig 2